\documentclass[sigconf]{acmart}
\AtBeginDocument{%
  }

\renewcommand\footnotetextcopyrightpermission[1]{}
\setcopyright{none} 
\usepackage{libertine}
\usepackage{algorithm}
\usepackage{algorithmic}
\usepackage{graphicx}
\usepackage{subcaption}
\usepackage{pifont}
\usepackage{color}
\usepackage{tikz}
\usepackage{multirow}
\usepackage{listings}
\usepackage{booktabs}

\newcounter{finding}

\newcommand{\finding}[2]{\refstepcounter{finding} \label{finding:#1}
  \begin{center}
  \begin{tikzpicture}%
    \node[rectangle, draw=black, top color=black!3, bottom
    color=black!3, rounded corners=2pt, inner xsep=5pt, inner
    ysep=6pt, outer ysep=10pt]{
    \begin{minipage}{0.95\columnwidth}
    \textbf{Answer to RQ\arabic{finding}}: #2
    \end{minipage}};%
  \end{tikzpicture}%
  \end{center}
}

\begin{document}

\title{Deep Reinforcement Learning for Automated Web GUI Testing}


\author{Zhiyu Gu}
\affiliation{%
  \institution{Institute of Software Chinese Academy of Sciences, University of Chinese Academy of Sciences}
  \city{Beijing}
  \country{China}
}
\email{guzhiyu22@otcaix.iscas.ac.cn}

\author{Chenxu Liu}
\affiliation{%
  \institution{Key Lab of HCST (PKU), MOE; SCS; Peking University}
  \city{Beijing}
  \country{China}
}
\email{chenxuliu@stu.pku.edu.cn}

\author{Guoquan Wu}
\authornote{Guoquan Wu is the corresponding author.}
\affiliation{%
  \institution{Institute of Software Chinese Academy of Sciences, University of Chinese Academy of Sciences}
  \city{Beijing}
  \country{China}
}
\email{gqwu@otcaix.iscas.ac.cn}

\author{Yifei Zhang}
\author{ChenXi Yang}
\affiliation{%
  \institution{Institute of Software Chinese Academy of Sciences, University of Chinese Academy of Sciences}
  \city{Beijing}
  \country{China}
}
\email{{zhangyifei, yangchenxi24}@otcaix.iscas.ac.cn}

\author{Zheheng Liang}
\affiliation{%
  \institution{Joint Laboratory on Cyberspace Security, China Southern Power Grid Guangdong Power Grid}
  \city{Guangzhou}
  \country{China}
}
\email{liangzheheng@xxzx.gd.csg.cn}

\author{Wei Chen}
\author{Jun Wei}
\affiliation{%
  \institution{Institute of Software Chinese Academy of Sciences, University of Chinese Academy of Sciences}
  \city{Beijing}
  \country{China}
}
\email{{wchen, wj}@otcaix.iscas.ac.cn}

\renewcommand{\shortauthors}{Gu et al.}

\begin{abstract}
Automated GUI testing of web applications has always been considered a challenging task considering their large state space and complex interaction logic. Deep Reinforcement Learning (DRL) is a recent extension of Reinforcement Learning (RL), which takes advantage of the powerful learning capabilities of neural networks, making it suitable for complex exploration space. In this paper, leveraging the capability of deep reinforcement learning, we propose WebRLED, an effective approach for automated GUI testing of complex web applications. WebRLED has the following characteristics: (1) a grid-based action value learning technique, which can improve the efficiency of state space exploration; (2) a novel action discriminator which can be trained during the exploration to identify more actions; (3) an adaptive, curiosity-driven reward model, which considers the novelty of an explored state within an episode and global history, and can guide exploration continuously.
We conduct a comprehensive evaluation of WebRLED on 12 open-source web applications and a field study of the top 50 most popular web applications in the world. The experimental results show that WebRLED achieves higher code/state coverage and failure detection rate compared to existing state-of-the-art (SOTA) techniques. Furthermore, WebRLED finds 695 unique failures in 50 real-world applications.
\end{abstract}

\maketitle

\section{Introduction}
Versatile web applications are playing an important role in a wide range of areas, such as online marketing, education, and news. However, the complex business logic that powers web applications with various functionality makes web testing increasingly challenging. In order to alleviate the labor cost of manual testing, automated web GUI testing approaches are proposed, which aim at maximizing code coverage and the number of triggered bugs in a limited time budget by interacting with the web applications under test using actions (e.g., click, drag) mimicking humans.

automated GUI testing approaches are widely studied, which can be classified into three types.
\textbf{Random-based} approaches are the most common strategies~\cite{monkey,mesbah2011invariant,white2019improving}, where pseudo-random operations are generated to fuzz web applications. However, they often generate invalid operations during exploration and also hardly explore some hard-to-reach states. 
\textbf{Model-based} approaches~\cite{mesbah2012crawling,su2017guided,yandrapally2022fragment} extract test cases from navigation models built by means of static or dynamic analysis. Guided by the model, they can access many web pages that can only be reached through the execution of long action sequences. Nonetheless, the automatically constructed model tends to be incomplete. In addition, the rapid evolution of web applications increases the complexity of maintaining these models. 
\textbf{Systematic strategies-based} approaches~\cite{biagiola2017search,biagiola2019diversity,wang2023parallel} use sophisticated techniques, such as evolutionary algorithms, to generate test inputs. However, they are more suitable for problems that have static environments and do not involve continuous decisions~\cite{romdhana2022deep}.

\begin{figure}[ht]
    \centering
    \includegraphics[width=\linewidth]{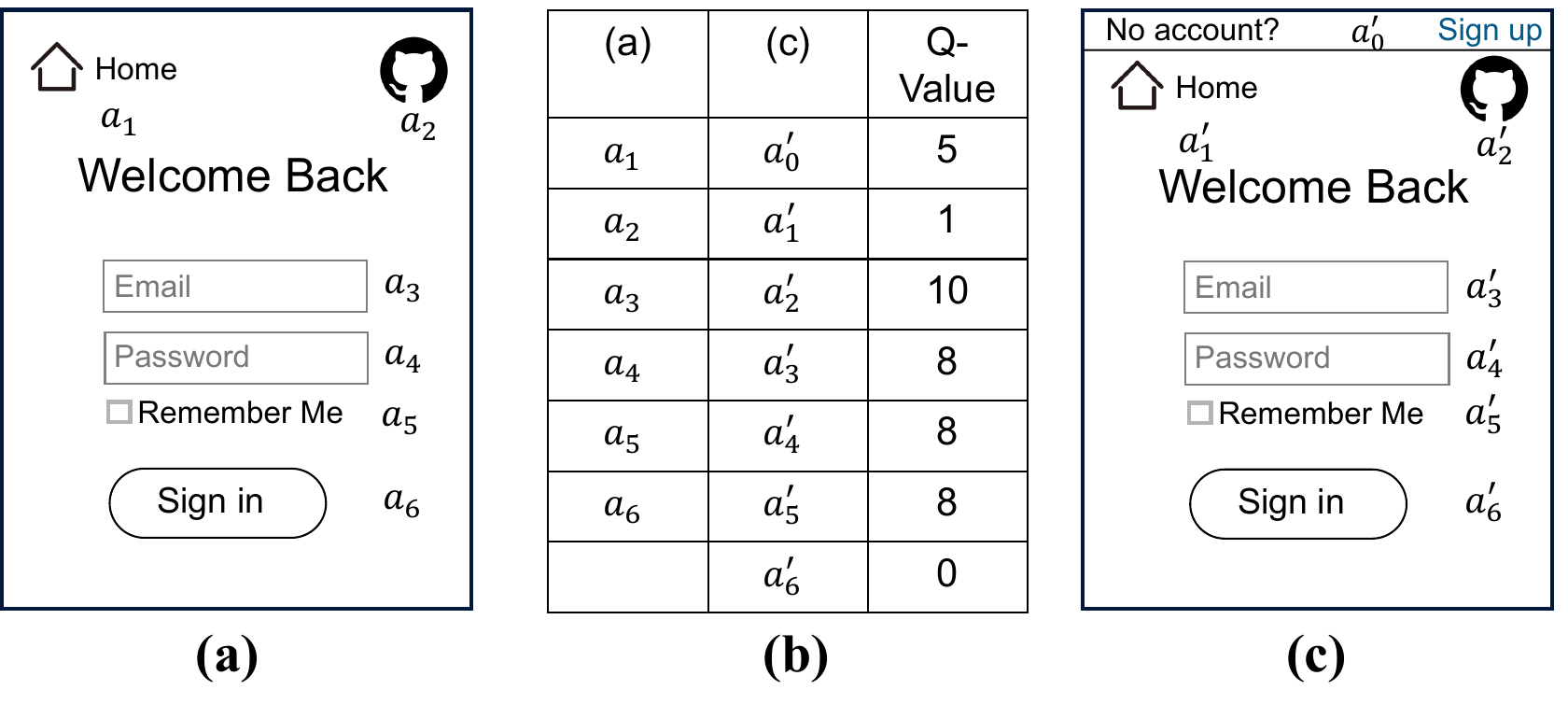}
    \caption{Example of list-based action space causing misalignment of actions.}
    \Description{Example of list-based action space causing misalignment of actions.}
    \label{fig:m_action_misalignment}
    \vspace{-15pt}
\end{figure}

Reinforcement Learning (RL) is a machine learning approach that enables an agent to learn an optimal policy to solve a task guided by the positive or negative reward retrieved from the environment, rather than by explicit supervision. RL has recently been applied to web testing~\cite{zheng2021automatic,sherin2023qexplore}. However, to date, only the basic form of RL (e.g., tabular RL~\cite{watkins1992q}) has been adopted, in which the value of each state-action pair is stored in a fixed table (called Q-Table). Deep Reinforcement Learning (DRL) is a recent extension of RL, which replaced tabular approaches with deep learning ones. The action value function is learned from the past experiences made by one or more Deep Neural Networks (DNNs), which is also called Deep Q-Networks (DQNs) in DRL. Depending on the powerful function approximation properties of DNN, DRL has proved to be substantially superior to tabular RL when the state space to explore is extremely large~\cite{romdhana2022deep}. Such capabilities make DRL suitable for exploring web applications with large state space. However, there are still some challenges that need to be addressed to unleash the capability of deep reinforcement learning to explore complex web applications effectively.

\textbf{Challenge 1: Appropriate representation of the action space for web testing}. Existing DRL-based applications~\cite{vinyals2019grandmaster,
jeong2020self,silver2018general,romdhana2022deep}, such as playing games and manipulating robotic arms, usually have the same action space for all states. In these studies, list-based action space representation is adopted to assign a sequentially generated number to represent each action. However, such representation is not suitable for automated web application testing, as each state has varying number of actions. Direct use of this technique will incur action misalignment between two similar states, making it hard to learn a stable action value function during exploration, and ultimately affect test efficiency.

Take a simple web application shown in Figure~\ref{fig:m_action_misalignment} as an example. Figure~\ref{fig:m_action_misalignment}.(a) and Figure~\ref{fig:m_action_misalignment}.(c) are two similar web pages, and the only difference is that Figure~\ref{fig:m_action_misalignment}.(c) has a hint at the top. To explore this application directly using existing DRL-based technique~\cite{liu2018reinforcement,jia2019dom}, the actions (represented by actionable web elements) on the page will be traversed and sequentially numbered. The action space can be seen as a list in which each index represents an action. During exploration, an action value function will be learned (represented by DQN), and for each page, it outputs a vector, which saves the value of the corresponding action in the action space. As the index of an action in the action space depends on the order of traversal, even small changes in the page can change its index in the action space, leading to the problem of action misalignment. Figure~\ref{fig:m_action_misalignment}.(b) shows the actions for page Figure~\ref{fig:m_action_misalignment}.(a) and Figure~\ref{fig:m_action_misalignment}.(c) and their corresponding values. It can be seen that the subtle difference between two pages results in the same actions not being aligned. The action with the highest value on page Figure~\ref{fig:m_action_misalignment}.(c) should be $a'_3$, but now it is $a'_2$.

The action misalignment problem can greatly slow the learning of the action value function. Reinforcement learning focuses on learning the mapping from states to action values. A stable mapping improves the DNN's ability to generalize across web pages and speeds up the learning process. However, the action misalignment problem disrupts this mapping and affects the stability of the predicted action value across similar states. It is possible to train DNN to distinguish similar states and learn the appropriate action value distribution for each state. However, it will take a long time to achieve this and is not suitable for web application testing with large state space.

\textbf{Challenge 2: General action recognition for web applications}. Unlike mobile applications, where actionable elements can be easily inferred based on the widget type, in web applications, each type of HTML element can be actionable. Existing work designs some heuristic rules to infer actionable elements based on the tag name and some attributes (e.g., $\textless input \textgreater$, $\textless button \textgreater$), and will miss lots of actionable elements. Ignoring these unobvious actions during the exploration will miss some important functionalities and will not test web applications adequately.

To address the aforementioned challenges, we propose WebRLED, an effective DRL-based approach for automated GUI testing of web applications. Specifically, \textbf{to alleviate action misalignment and explore web applications efficiently}, a novel grid-based action space representation and action value learning technique is proposed, which divides the web page into a grid consisting of multiple cells. During exploration, DQN is trained to first estimate the value of each cell on a page, and then upsampling technique~\cite{wang2020deep} is utilized to calculate the value of each actionable element based on the value of its surrounding cells. \textbf{To recognize more actionable elements during exploration}, in addition to some common heuristic rules, WebRLED also trains a novel action discriminator by trying to trigger more actionable elements, which can assist DRL-based agent to test more functionalities of the application. Moreover, we also design an adaptive, curiosity-driven reward model, which considers the novelty of an explored state within an episode and global history and can guide WebRLED to explore the application continuously to reach some deep states.

To evaluate the effectiveness of WebRLED, we perform a comprehensive evaluation on a total of 12 open-source web applications. Experiments show that WebRLED achieves higher code/state coverage and failure detection rate compared to existing work. It also finds 695 unique failures in 50 real-world applications.

Generally, this paper makes the following contributions:
\begin{itemize}
    \item To the best of our knowledge, WebRLED is the first publicly available DRL-based approach for automated GUI testing of web applications.
    \item To effectively and efficiently explore web applications, (1) we propose a grid-based action value learning technique, which combines DQN and upsampling technique to approximate the best action value in a given state; (2) we design a new action discriminator, which is trained during exploration and can recognize more actionable web elements; (3) we design an adaptive reward model that takes into account the novelty of a state within an episode and the overall exploration history.
    \item We implement a tool and perform a comprehensive evaluation. The results show that our approach outperforms existing ones in terms of code/state coverage and failure detection. The artifact is available on an open-source link~\cite{webrled}.
\end{itemize}

\section{Background} \label{background}
\subsection{Reinforcement Learning}
Reinforcement learning is a group of machine learning algorithms that train an agent to learn strategies by interacting with the environment to achieve specific goals. It can be formalized as a Markov decision process (MDP), which is defined by a tuple $\langle\mathcal{S},\mathcal{A},\mathcal{R},\mathcal{P} \rangle$ consisting of a state space $\mathcal{S}$, an action space $\mathcal{A}$, a reward function $\mathcal{R}:\mathcal{S}\times\mathcal{A}\rightarrow\mathbb{R}$ with $r_t=R(s_t,a_t,s_{t+1})$ and a transition probability function $\mathcal{P}$.
$\mathcal{P}(s_{t+1} | s_t, a_t)$ represents the probability of transitioning from state \(s_t\) to state \(s_{t+1}\) while taking action \(a_t\).

The goal in RL is to learn a policy $\pi$ that maximizes the expected return, which is calculated as: $R= \sum^\infty_{t=0} \gamma^{t}r_t$. A discount factor $\gamma \in (0,1)$ is needed for convergence, determining how much the agent cares about the rewards in the distant future relative to those in the immediate future. $\tau=(s_1, a_1, r_1, s_2, ...)$ is a sequence of states and actions in the environment, named \textit{episode}. Testing a web application can be seen as a task divided into finite-length episodes.

Reinforcement learning usually involves learning Q value functions to estimate how good it is to perform an action in a state. Traditional RL, such as Q-Learning, uses tables (called Q-Table) to represent the current estimate of the action value function $Q(s,a)$. In state $s_t$, when an action $a_t$ is taken, the associated state-action value in the Q-Table is updated using the formula:
\begin{equation}
\resizebox{0.9\linewidth}{!}{$Q(s_t,a_t) = Q(s_t,a_t)+\alpha(r_t+\gamma\max_aQ(s_{t+1},a)-Q(s_t,a_t))$
}
\end{equation}
where $\alpha$ is the learning rate and $\max_aQ(s_{t+1},a)$ gives the maximum value for future rewards across all actions.

However, tabular RL struggles with high-dimensional problems~\cite{lillicrap2015continuous}. For large discrete state/action spaces, representing all states and actions in a Q-table becomes impractical, leading to instability and difficulty in learning optimal policies. DRL is a recent extension to tabular RL. Relying on the powerful function approximation properties of deep neural network, it provides new ways to overcome the limitation of tabular RL. One of the earliest DRL algorithms is Deep Q-Network (DQN)~\cite{mnih2013playing}, which uses convolutional neural networks to approximate the action value function. With continuous improvements to DQN (e.g., R2D2~\cite{kapturowski2018recurrent}), DRL now handles tasks with long-term dependencies more effectively.

\subsection{Upsampling}
\begin{figure}[ht]
    \centering
    \includegraphics[width=\linewidth]{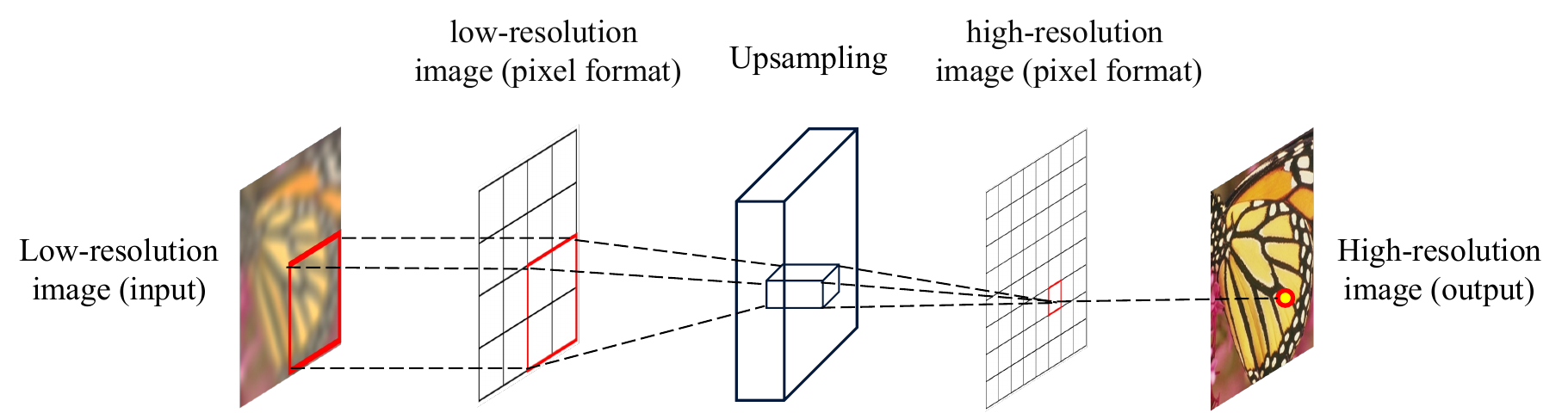}
    \caption{Upsampling for enhanced image resolution.}
    \Description{Upsampling for enhanced image resolution.}
    \label{fig:upsampling}
\end{figure}

In the area of image processing, upsampling~\cite{kopf2007joint} is a widely used technique to enhance the resolution of the image. Upsampling increases the size of the original image and adopts interpolation algorithms to fill in the newly added areas. As shown in Figure~\ref{fig:upsampling}, the low-resolution image can be transformed into a large high-resolution image.  The value of each new added pixel is determined by synthesizing the values of the surrounding pixels from the low-resolution image. Inspired by the idea of upsampling technique, in this work, we apply it to better distinguish the action value for the web elements that are close to each other on the page.

\section{Approach} \label{approach}

\subsection{An Overview of WebRLED}

\begin{figure*}[ht]
    \centering
    \includegraphics[width=\linewidth]{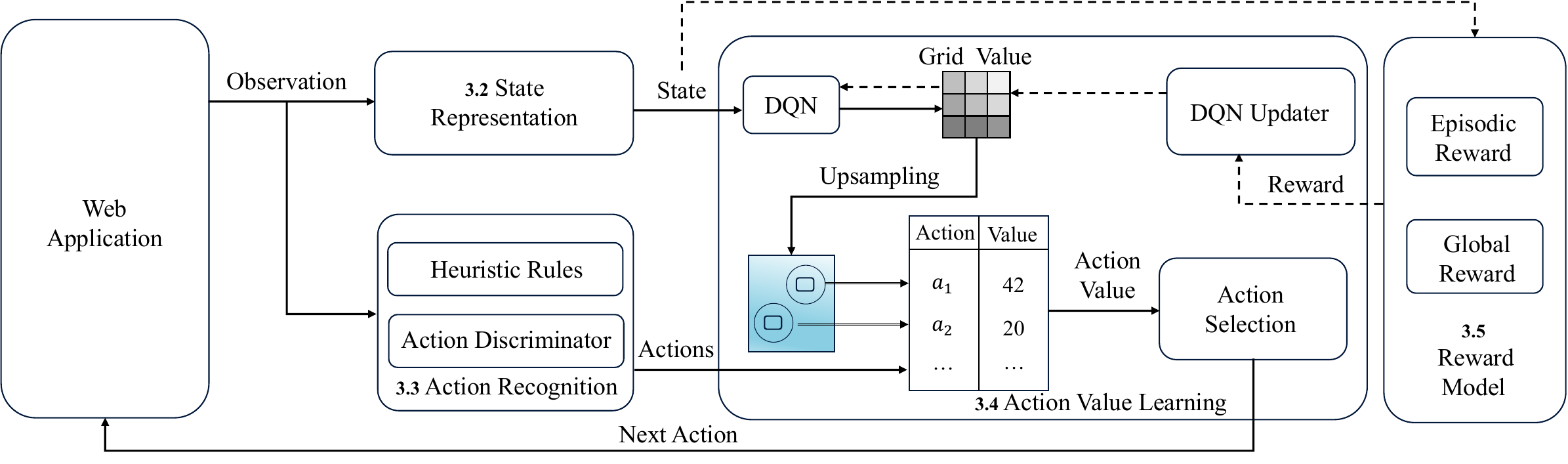}
    \caption{The Workflow of WebRLED}
    \label{fig:a_overview}
    \vspace{-15pt}
\end{figure*}

WebRLED is an automated end-to-end web testing tool based on deep reinforcement learning. It continuously updates its policy during exploration, generates action sequences, and interacts with web applications under the guidance of rewards. The goal of WebRLED is to explore as many states as possible as well as to discover the underlying defects in web applications. Figure~\ref{fig:a_overview} shows the overall approach of WebRLED. \ding{182} To apply DRL for web testing, the state representation module first abstracts the observation into state, where an observation is a raw HTML document, and the state is the semantic representation (the embedding) of the page. \ding{183} The action recognition module identifies actionable elements on the current page as actions. Besides the common heuristic rules for identifying actions, during exploration, we also train an online action discriminator based on multi-layer perceptron (MLP) to recognize more actionable elements. \ding{184} The action value learning module combines the DQN and upsampling technique to predict the value for each action on the current page. \ding{185} To train the DQN, action selection module applies the $\epsilon$-greedy strategy to select an action, and perform it to try to expose new state. \ding{186} The reward model considers the novelty of the new state within an episode and global history, and computes a reward for the selected action. \ding{187} Based on the reward received, the DQN updater module calculates the rewards of the cells surrounding the selected action based on their distance to this action on the page. In the following, we will introduce the key techniques in WebRLED, including state representation, action recognition, action value learning, the designed reward model, and the whole exploration strategy.

\subsection{State Representation} \label{state_representation}

To apply DRL-based exploration, our approach first generates embedding vector representations for each explored web page. We use the state abstraction function WebEmbed~\cite{stocco2023neural}, which leverages neural network embeddings and threshold-free classifiers to accurately represent and detect near-duplicate pages. Doc2Vec~\cite{2014Distributed}, a popular document embedding technique, processes HTML pages containing both tags and text. WebEmbed is pre-trained on an unlabeled corpus of web pages using the Doc2Vec model, making it ideal for computing the semantic representation of HTML pages. However, for complex web applications, long HTML documents can introduce noise in the embeddings, making it difficult for the agent to recognize states. To address this, we simplify the original HTML document using two steps.

\begin{itemize}
    \item Step1: Remove elements by tag names. The elements in the $\textless head \textgreater$ part, such as $\textless link \textgreater$ and $\textless style \textgreater$, are removed as they will not be shown directly on the page. For the same reason, the $\textless script \textgreater$ elements that appears in the $\textless body \textgreater$ part are also removed. 
    \item Step 2: Remove duplicates. If all children of a DOM element have the exact same structure, we keep only the first child and remove the rest. We then mask the text content of the remaining child nodes. Such a simplification can identify similar web pages. Additionally, we use the number of removed nodes as an attribute value for the remaining child node to represent subtle differences between similar pages.
\end{itemize}

The simplified HTML document removes redundant elements. It also highlights common features while preserving differences between similar states, which helps to speed up the learning of DQN. Finally, WebEmbed embeds the simplified HTML into a one-dimensional vector, making it suitable for DRL-based exploration.

\subsection{Action Recognition}  \label{action_recognition}
To recognize more actions on the page, we propose an action recognition technique combining heuristic rules and an online trained action discriminator. During exploration, WebRLED first applies heuristic rules to identify actionable elements. Table~\ref{tab:heuristic_rules} shows the action type supported by default according to the tag name. Note that, for $\textless input \textgreater$ tag, its action type can be further inferred based on its attribute \textit{type}. If the value is ``button'', the action type is \textit{click}, and if the value is ``text'', the action type is \textit{input}. For the \textit{input} action, the value is generated using a combination of manual configuration and random generation currently, as detailed in Section~\ref{experiment_setup} (3). 

\begin{table}[ht]
    \centering
    \caption{Heuristic rules for action recognition.}
    \resizebox{0.95\linewidth}{!}{
    \begin{tabular}{c|ccccccc}
        \hline
        \multirow{2}{*}{Action type} & \multicolumn{7}{c}{Tag Name of DOM Element}\\
        \cline{2-8}
        & a & button & input & textarea & form & fieldset & select  \\
        \hline
        click       & $\surd$ & $\surd$ & $\surd$ &  &  &  & \\
        input       &  &  & $\surd$ & $\surd$ &  &  & \\
        form-fill   &  &  &  &  & $\surd$ & $\surd$ & \\
        select      &  &  &  &  &  & & $\surd$ \\
        \hline
    \end{tabular}}
    \label{tab:heuristic_rules}
\end{table}

As many actionable web elements cannot be recognized just based on heuristic rules, WebRLED also trains an action discriminator during exploration. The action discriminator is a multi-layer perceptron (MLP) designed as a four-layer structure for efficiency and performance, using the Rectified Linear Unit (ReLU) activation function to enhance nonlinearity. Its input is the semantic representation of the DOM element, obtained by encoding the attributes of the element using Bert, and the output is an integer indicating the action type. The action discriminator now supports recognizing the action type \textit{click} (denoted as 1) and \textit{dbclick} (denoted as 2). It can be extended to contain more types of action.

The training data for the action discriminator come from two sources: specialized data collection at the end of each episode and feedback from action within an episode. First, after completing an episode, all leaf nodes on the page that are not recognized as actions by the heuristic rules are executed in turn, and the attributes of the target element and the performed action are collected (If the page does not change, the action type is 0) to train the action discriminator. After training, for a new episode, WebRLED will use it to predict the actionable elements on the page, and add them into the action space of the current page. Depending on the result after performing an action, the corresponding data will be collected to retrain the action discriminator to make it more accurate as the exploration progresses.

\subsection{Action Value Learning}  \label{action_value_learning_and_estimation}
To address test inefficiency caused by the action misalignment mentioned in the introduction, we propose a grid-based action value learning approach. Unlike existing work, which trains a DQN (action value function) to directly predict the action value, WebRLED divides a web page into a grid consisting of N$\times$N cells, where N are hyperparameters chosen on the basis of the experiment, and then trains a DQN to predict the value of each cell in the grid. The cell value represents the cumulative reward when performing actions in this area. A higher value of the cell represents a higher reward for choosing actions around it. 

Based on the estimated cell values by DQN, WebRLED continues to calculate the value of actionable element. Note that we do not use the cell value to represent the value of the action contained in the cell considering the following situations: (1) one cell may contain multiple actionable elements; (2) some actionable elements may cross multiple cells. In both cases, it is hard to select the action just using the cell value.  

To better distinguish the value of different actions and provide a consistent and relatively fair value for each action on the page, inspired by the upsampling technique, WebRLED calculates an action's value based on the values of surrounding cells. Specifically, the value of each action in the action space is determined by summing the values of cells within a circle centered on this action. A cell is considered covered if the distance from its center to the action's center is less than or equal to the radius of the circle. Currently, the radius of the circle is empirically set to 1.5 times the cell's length. The reasons are as follows: (1) if the radius is less than the cell's length, upsampling technique will not work to better distinguish the value of each action on the page; (2) large radius will increase computational costs and also require more time to learn a successful policy, and thus affect the exploration efficiency.

Based on the estimated value for each action on the page, WebRLED applies $\epsilon$-greedy algorithm to select an action. After performing the action, a reward will be calculated to update DQN based on the designed reward model. Note that DQN trained in WebRLED represents the approximation of the grid value function, not the action value function. To use the action reward to update the DQN, Q-value of each cell surrounding this action will be updated according to Formula~\ref{Our learning formula}, in which $\beta_i$ represents the contribution of each cell to the action reward, which is calculated as the ratio of its distance to the selected action to the radius. The closer the cell is to the action, the more contribution it makes to the action selection and the higher reward it will receive.

\begin{equation}
\begin{aligned}
Q(s_t, a_t^1) &\leftarrow \beta_1 r_t + \lambda \max_{a_{t+1}'} Q(s_{t+1}, a_{t+1}') \\
Q(s_t, a_t^2) &\leftarrow \beta_2 r_t + \lambda \max_{a_{t+1}'} Q(s_{t+1}, a_{t+1}') \\
& \cdots \\
Q(s_t, a_t^n) &\leftarrow \beta_n r_t + \lambda \max_{a_{t+1}'} Q(s_{t+1}, a_{t+1}') \\
\beta_i &= \frac{\sqrt{{(c_x - p^i_x)}^2 + {(c_y - p^i_y)}^2}}{R}
\end{aligned}
\label{Our learning formula}
\end{equation}

\textbf{Discussion}. Compared to list-based action space representation, grid-based technique aligns the actions and corresponding values according to their locations on the page. During exploration, the action values learned from previous page can be applied to the new page if the embedding of these two pages are similar, which improves the learning efficiency of DQN. For the different part in the new page, even if the action value may not be the true value, it still can be updated later by DQN based on new collected training data. However, for the list-based technique, the mapping of the same actions on similar pages is chaotic, significantly increasing the learning difficulty.

\subsection{Reward Model}
\begin{figure}[ht]
    \centering
    \includegraphics[width=0.95\linewidth]{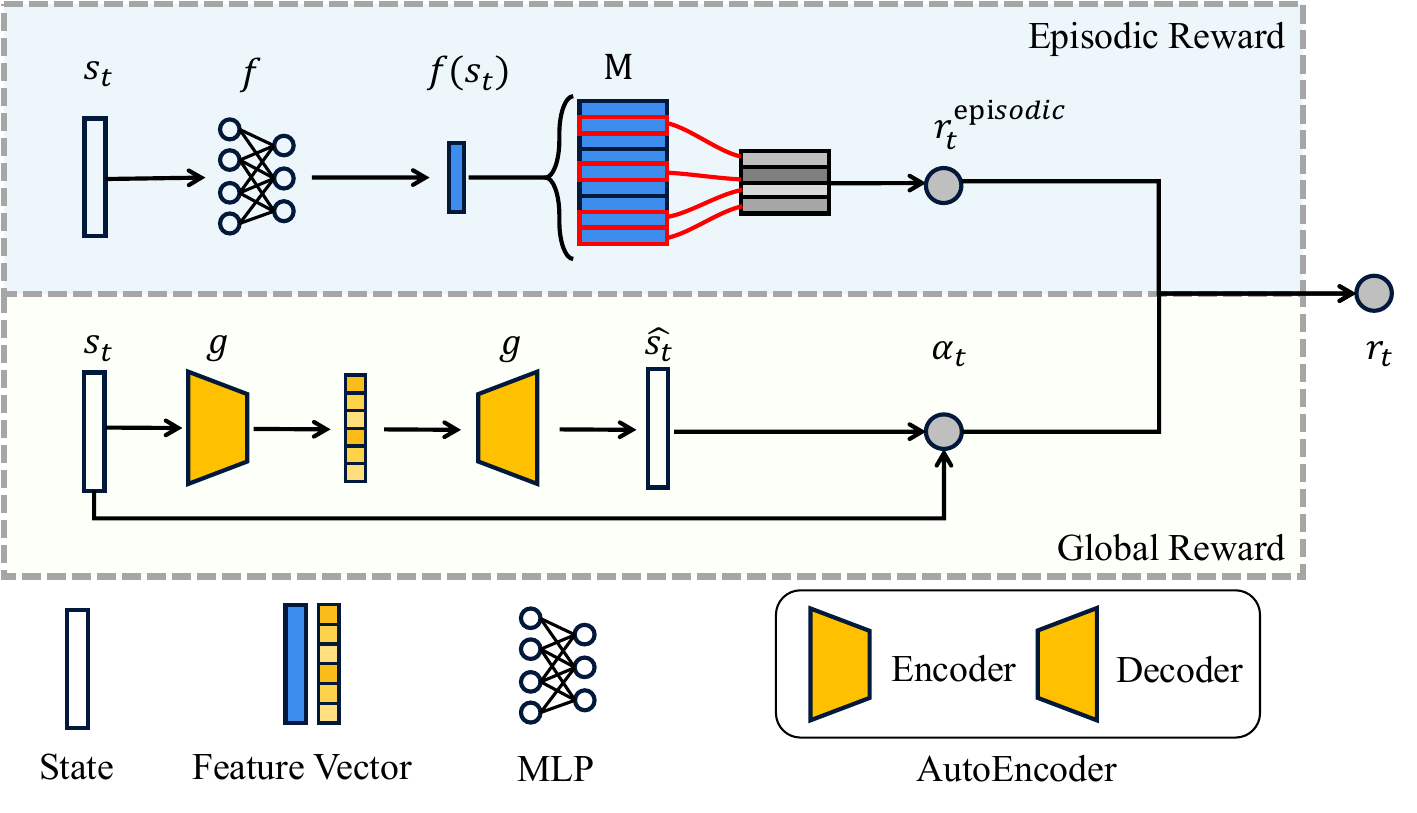}
    \caption{The reward model of WebRLED.}
    \Description{The reward model of WebRLED.}
    \label{fig:reward_model}
    \vspace{-20pt}
\end{figure}

As shown in Figure~\ref{fig:reward_model}, the reward model in WebRLED consists of two parts: episodic reward and global reward. The reward models of the existing work~\cite{zheng2021automatic,pan2020reinforcement,romdhana2022deep} mainly consider the novelty of an explored state in the global exploration history. The rewards decrease as the number of visits to that state increases and cannot guide the agent to continually explore the application with complex state space. To address this limitation, the episodic reward introduced in WebRLED is calculated based on the number of times a state has been visited in the current episode, which can still provide dense rewards in the later stages of exploration. Specifically, the episodic reward model in WebRLED contains an MLP-based feature extraction network and an episode buffer $M$. The network encodes the state $s_t$ to a feature vector ${f}_t$ to identify similar states. The buffer $M$ stores the feature vectors of all the states visited in the current episode and is reset at the beginning of each episode. The episodic reward is calculated as $r^{episodic}_t = \frac{1}{\sqrt{n({f}_t)}}$, where $n({f}_t)$ is the count of feature vectors in the buffer $M$ that is most similar to ${f}_t$.

However, only using episodic reward can cause the agent to visit some states frequently without an incentive to explore new states. Inspired by NGU~\cite{badia2020agent57,badia2020never}, the reward model is calculated by multiplicatively modulating the episode reward with a global novelty factor $\alpha_t$, shown as follows: $r_t = r^{episodic}_t \cdot min\{max\{\alpha_t, 1\}, L\}$

where L is a chosen maximum reward scaling (see Section~\ref{experiment_setup}(3)). Mixing rewards in this way, we leverage the long-term novelty that $\alpha_t$ offers, while $r_t$ consistently offers rewards and directs the agent towards underexplored states.

Our global reward model uses an autoencoder to calculate rewards based on the number of times the state is visited globally. An autoencoder is a type of neural network architecture designed to efficiently compress input data down to its essential features, then reconstruct the original input from this compressed representation. The autoencoder has smaller reconstruction errors for similar data appearing in the history and larger errors for new data. Thus, we can assess the novelty of the state based on the reconstruction error. The larger the value of the reconstruction error, the more novel the state is. The autocoder is updated at the end of the episode based on all states visited during the episode, so it contains the global exploration history. Specifically, the formula of the global factor is $\alpha_t = ||g(s_t;\theta) - s_t||^2$, where $g$ is an autoencoder and $\theta$ represents its parameters.

\subsection{Exploration Strategy}

Algorithm~\ref{algo_webrled} describes the exploration process of WebRLED, which takes a target web application $env$, a maximum number of episodes $n$, a maximum number of steps per episode $m$, and exploration probability $\epsilon$ as inputs. It outputs a set of action sequences $S$. Firstly, WebRLED initializes the DQN $\pi$, action discriminator $D$, reward model $R$ (Line 1), variables $phase$, $trajectories$, $T$ and $S$ (Line 2). Here, $phase$ is a boolean value indicating the current phase of action recognition, $trajectories$ are the data collected during episodes to update the DQN, and $T$ is the training data to update $D$. The process then tests the web application until it completes $n$ episodes (Line 3). At the beginning of each episode, WebRLED creates a new list $traj$ to collect experiences and resets $env$ to its home page $o$ (Lines 4-5). It performs up to $m$ steps in each episode (Line 6). At the beginning of each step, it abstracts the current page $o$ into a state $s$ and recognizes actions based on the $phase$ value (Line 7-8), 0 means that only heuristic rules are applied, while 1 means that $D$ also works. The action value is then evaluated using DQN and upsampling technique (Line 9). After that, $\epsilon-greedy$ algorithm is applied to select an action (Lines 10), which is executed on $env$, leading to a new page $o'$ (Line 11). The action $a$ will be added to action sequence $A$ (Line 12). If $phase$ is 1, we will update the training data $T$ based on the feedback from the action execution (Line 13-15). WebRLED then abstracts $o'$ into a new state $s'$ to calculate the reward $r$ (Lines 16-17). Then WebRLED stores the tuple $(s, a, s', r^{total})$ as an experience in $traij$ and updates $o$ to $o'$ (Lines 18-19). At the end of episode, experiences are added to $trajectories$ (Line 21). The reward model $R$ is updated using $traj$, while $trajectories$ is used to update the DQN (Lines 22-23). WebRLED then tries to execute the elements of the page that are not recognized as actions by the heuristic rules and collects the execution results to update $D$ (Line 24). If $phase$ is 0 and the length of $D$ exceeds the threshold $\zeta$, then $phase$ is set to 1, and the action discriminator $D$ is trained with the data $T$ (Lines 25-28). Otherwise, it's updated every ten episodes (Lines 31). Finally, WebRLED returns $S$, which saves the explored action sequences (Line 34).

\begin{algorithm}
    \renewcommand{\algorithmicrequire}{\textbf{Input:}}
    \renewcommand{\algorithmicensure}{\textbf{Output:}}
    \caption{WebRLED}
    \label{algo_webrled}
    \begin{algorithmic}[1]
    \REQUIRE{The target web application \( env \), \( n \), \( m \), \( \epsilon \)}
    \ENSURE{Set of action sequences \( S \)}
    \STATE Initialize \( \pi \), action discriminator \( D \), reward model \( R \)
    \STATE \( phase \gets 0, trajectories \gets [], T \gets [], S \gets \{\} \)
    \FOR{\( i \leftarrow 0 \) \textbf{to} \( n \)}
        \STATE \( traj \gets [], A \gets [] \)
        \STATE \( o \gets \text{reset}(env) \)
        \FOR{\( j \leftarrow 0 \) \textbf{to} \( m \)}
            \STATE \( s \gets \text{stateRepresentation}(o) \)
            \STATE \( actions \gets \text{actionRecognition}(o, phase, D) \)
            \STATE \( action\_value \gets \text{actionEstimation}(s, actions, \pi) \)
            \STATE \( a \gets \text{actionSelect}(\epsilon, actions, action\_value) \)
            \STATE \( o' \gets \text{envStep}(a) \)
            \STATE \( \text{append}(A, a) \)
            \IF{\( phase \) is 1}
                \STATE \( \text{updateData}(o, a, o', T) \)
            \ENDIF
            \STATE \( s' \gets \text{stateRepresentation}(o') \)
            \STATE \( r \gets \text{rewardCalculation}(s, a, s', R) \)
            \STATE \( traj \gets traj \cup \{(s, a, s', r)\} \)
            \STATE \( o \gets o' \)
        \ENDFOR
        \STATE \( \text{append}(trajectories, traj) \)
        \STATE \( \text{updateRewardModel}(traj, R) \)
        \STATE \( \text{updateDQN}(trajectories, \pi ) \)
        \STATE \( \text{collectData}(env, T); S.\text{add}(A) \)
        \IF{\( phase \) is \( 0 \)}
            \IF{\( \text{length}(T) > \zeta \)}
                \STATE \( phase \gets 1 \)
                \STATE \( \text{updateActionDiscriminator}(D, T) \)
            \ENDIF
        \ELSE
            \STATE \( \text{updateActionDiscriminator}(D, T, i) \)
        \ENDIF
    \ENDFOR
    \RETURN \( S \)
\end{algorithmic}
\end{algorithm}

\section{Evaluation} \label{evaluation}
We implement WebRLED based on Python 3.10.14 and Pytorch 2.3.0~\cite{paszke2019pytorch} with more than 5000 lines of code. We build a reinforcement learning environment for web applications based on the OpenAI Gym interface~\cite{1606.01540}, which is a de-facto standard in the RL field. The implementation of DQN is based on R2D2~\cite{kapturowski2018recurrent}. 

To demonstrate the effectiveness of WebRLED, we answer the following four research questions.

\begin{itemize}
    \item \textbf{RQ1 (Code Coverage):} How is WebRLED's exploration capability in terms of code coverage?
    \item \textbf{RQ2 (Failure Detection):} How effective is WebRLED in detecting web applications failures?
    \item \textbf{RQ3 (Ablation Experiment):} How do the proposed techniques in WebRLED, including grid-based action value learning, action discriminator and episodic reward model, improve test efficiency and effectiveness? 
    \item \textbf{RQ4 (Scalability):} How effective is WebRLED in testing real-world web applications?
\end{itemize}

\subsection{Experiment Setup} \label{experiment_setup}
\begin{table*}
    \centering
    \caption{Experimental subjects. (For applications without a labeled version, we use time as a substitute for the version.)}
    \resizebox{\linewidth}{!}{
    \begin{tabular}{c|cccc||c|cccc}
        \hline
        Web apps   & Version & Client LOC & Server LOC & Description &
        Complex Web apps   & Version & Client LOC & Server LOC & Description \\
        \hline
        Dimeshift~\cite{Dimeshift}  & 2018   & 5,140 (JS) & 3,298 (JS) & Finance & 
        Timeoff~\cite{timeoff} & 1.0.0 & 2,937 (Handlebars) & 7,933 (JS) & Attendance \\

        Pagekit~\cite{Pagekit} & 1.0.15 & 4,214 (JS) & 13,856 (PHP) & Publishing &
        Realworld~\cite{realworld} & 2024 & 7,604 (JS) & 2,705 (TS) & Blog \\

        Splittypie~\cite{Splittypie} & 2018 & 2,710 (JS) & 829 (JS) & Finance &
        4gaBoards~\cite{4gaBoards} & 3.1.9 & 16,655 (JS) & 10,450 (JS) & Collaboration \\
        
        Phoenix~\cite{Phoenix}    & 1.1    & 2289 (JS) & 1,135 (Elixir) & Management &
        Parabank~\cite{parabank}  & 2024 & 2,662 (JSP) & 9,446 (Java) & Finance \\ 
        
        Retroboard~\cite{Retroboard} & 2018 & 2144 (JS) & 278 (JS) & Collaboration &
        Gadael~\cite{gadael} & 0.1.4 & 6,265 (JS) & 7,811 (Java) & Management \\

        Petclinic~\cite{Petclinic}  & 2018   & 2,939 (JS) & 842 (Java) & Healthcare &
        Agilefant~\cite{agilenfant}  & 3.5.4 & 3,949 (JSP) & 27,584 (Java) & Management \\
        \hline
    \end{tabular}
    }
    \label{tab:webapps_info}
\end{table*}

(1) \textbf{Benchmarks:} To perform a comprehensive evaluation, three benchmarks were used in the experiments. The first benchmark includes 6 applications widely used by existing work~\cite{zheng2021automatic,sherin2023qexplore,chang2023reinforcement}. 
To evaluate the effectiveness of WebRLED on complexed web applications, we further construct the second benchmark, which consists of 6 complex modern web applications collected from recent work~\cite{yandrapally2023carving,chang2023reinforcement} and Github. The third benchmark contains the top 50 popular web applications in the world~\cite{top50} to evaluate the effectiveness in real usage. 
Table~\ref{tab:webapps_info} shows the 12 open-source web applications contained in the first two benchmarks. The LOC column shows the lines of code in the client and the server for each selected web application. From the LOC's point of view, the applications in the second benchmark are more complex than those in the first benchmark. Each application in our second benchmark exceeds 10,000 LOC. This standard is also used by Do{\u{g}}an et al.~\cite{dougan2014web} as an indicator to distinguish between toy and real-world applications.

(2) \textbf{Baseline Approaches:} To evaluate the effectiveness of WebRLED, we choose four state-of-the-art approaches, including WebExplor~\cite{zheng2021automatic},  Crawljax~\cite{mesbah2012crawling}, FEEDEX~\cite{fard2013feedback}, and FragGen~\cite{yandrapally2022fragment}, as baselines for comparison.

\begin{itemize}
    \item WebExplor~\cite{zheng2021automatic} employs curiosity-driven reinforcement learning for generating test cases and builds an automaton for enhancing test efficiency.
    \item Crawljax~\cite{mesbah2012crawling} is a model-based testing tool for modern web applications, which dynamically builds a state transition graph during exploration. 
    \item FEEDEX~\cite{fard2013feedback} is a feedback-directed web application exploration tool. It optimizes test model generation considering code coverage, structural diversity, etc.
    \item FragGen~\cite{yandrapally2022fragment} uses fragment-based state abstraction and fine-grained analysis to enhance the state abstraction in Crawljax, and to further benefit the web application exploration effectiveness.
\end{itemize}

(3) \textbf{Configurations:} In all experiments, each tool is allocated the same time budget of 1 hour. To ensure fairness, identical settings are applied to all tools, including the configuration of login scripts and the disabling of specific actions, such as external links and ``logout'' button. Additionally, both click and page refresh waiting times are set to 2000 ms for each tool. These configurations are uniformly applied across all subjects. To counteract randomness from a statistical perspective, we repeated each experiment five times and calculated the average result.

For Crawljax, FEEDEX, and FragGen, the form-filling mode is configured to random, with elements being clicked in a random sequence. Both the crawl depth and the number of states are set to infinite. For WebExplor, the parameters associated with the Q-Learning algorithm are as follows: $\gamma$ is set to 0.65, the learning rate is set to 0.95, and the maximum number of steps per episode is set to 100, along with other configurations. For WebRLED, the hyperparameters are set as epsilon = 0.4, L = 5, and $\gamma$ = 0.95, based on previous work~\cite{badia2020agent57,badia2020never}. Note that, The input value for WebRLED is generated randomly according to the input type. However, some input types, such as ``username'' and ``password'', have fixed corresponding generated values that are predefined in the configuration file, ensuring consistency with the input settings of other tools.

(4) \textbf{Metric:} We use code coverage and the number of failures as metrics to evaluate the effectiveness of tools.
\begin{itemize}
    \item \textbf{Coverage:} For fairness in comparison, client-side code coverage is calculated for the first benchmark to match the settings used in previous work~\cite{zheng2021automatic,sherin2023qexplore,chang2023reinforcement}. For the second benchmark, server-side code coverage is chosen as the logic in the backend is more complex compared to the client. Nyc~\cite{nyc} and JoCoCo~\cite{jacoco} are used to calculate the coverage for JavaScript and Java code, respectively.
    
    \item \textbf{Failure:} Similar to WebExplor, in WebRLED, failure is defined as system-level log errors reported in the browser's console. These failures can be captured by testing tools using Selenium or Playwright. The number of failures is manually deduplicated by identifying those that are essentially the same but differ only in parameters. For example, in the following code snippet, each line represents a failure, with identical content omitted and differences highlighted in bold. We consider line 1 and line 2 as distinct failures because their different parameters correspond to different objects and functions. However, line 3 and line 4, despite having different parameters, represent the same failure.
\end{itemize}

\begin{lstlisting}
  Error: Param values not valid for state ``petNew''
  Error: Param values not valid for state ``ownerEdit''
  Access ... at ``http://gravatar.com/avatar/?r=g&s=560&d=blank''
  Access ... at ``http://gravatar.com/avatar/?r=g&s=80&d=blank''
\end{lstlisting}

\subsection{Experiment Design} \label{experiment_design}

\textbf{For RQ1,} we evaluate the exploratory effectiveness of WebRLED by comparing it with four SOTA approaches. The average code coverage achieved serves as a metric to assess the exploratory capability, as detailed in Section~\ref{experiment_setup} (4). \textbf{For RQ2,} we investigate the failure detection capability of WebRLED and the baseline approaches using failures, as defined in Section~\ref{experiment_setup} (4). \textbf{For RQ3,} we conduct ablation experiments comparing WebRLED with three variants: \textit{WebRLED+}, \textit{WebRLED-}, and \textit{WebRLED*}. \textit{WebRLED+} utilizes a list-based action space representation, \textit{WebRLED-} omits the training of an action discriminator for action identification, and \textit{WebRLED*} focuses solely on the global reward by excluding the episodic reward. The evaluation involved six subjects: three from the first benchmark and three from the second. \textbf{For RQ4,} we evaluate the effectiveness of WebRLED in real scenarios using the world's 50 most popular applications and analyze the failures discovered.

\subsection{Code Coverage (RQ1)}

\begin{table*}[htbp]
    \centering
    \caption{Comparison of relevant baselines for average branch coverage and failure detection in first benchmark.}
    \resizebox{\textwidth}{!}{
    \begin{tabular}{c||ccccc||ccccc}
        \hline
        \multirow{2}*{Web apps} & \multicolumn{5}{c||}{Average Branch Coverage (\%)}  
                       & \multicolumn{5}{c}{Average number of failures (\#)}  \\
        \cline{2-11}
                      & Crawljax  & FEEDEX & FragGen & WebExplor & WebRELD  
                      & Crawljax  & FEEDEX & FragGen & WebExplor & WebRELD \\
        \hline
        Dimeshift  & 26.89 \% & 35.61 \% & 29.91 \% & 54.22 \%   & \textbf{55.37} \%  
                      & 4.6 & 3.8 & 4.6 & 3.0 & \textbf{5.4} \\
                      
        Pagekit    & 29.13 \% & 22.81 \% & 37.89 \% & 25.79 \%   & \textbf{39.76} \%  
                      & 1.6 & 5.8 & 4.6 & 6.6 & \textbf{7.4} \\
                      
        Splittypie & 41.71 \% & 12.22 \% & 43.76 \% & 36.16 \%   & \textbf{44.61} \%  
                      & \textbf{5.0} & 4.0 & 4.4 & 4.0 & \textbf{5.0} \\
                      
        Phoenix    & 69.74 \% & 48.68 \% & 62.37 \% & 71.84 \%   & \textbf{82.89} \%  
                      & 0.0 & 0.8 & 0.0 & 0.0 & \textbf{2.0} \\
                      
        Retroboard & 53.80 \% & 49.00 \% & 55.56 \% & 60.23 \%   & \textbf{77.43} \%  
                      & \textbf{1.0} & 0.6 & \textbf{1.0} & \textbf{1.0} & \textbf{1.0} \\
        
        Petclinic  & 70.00 \% & 20.00 \% & 77.00 \% & \textbf{85.00} \%   & \textbf{85.00} \%  
                      & 1.6 & 1.2 & 0.8 & \textbf{3.8} & \textbf{3.8} \\
        \hline
        Average    & 48.55 \% & 31.39 \% & 51.08 \% & 55.54 \% & \textbf{64.18} \% 
                      & 2.3 & 2.7 & 2.6 & 3.1 & \textbf{4.1} \\
        \hline
    \end{tabular}}
    \label{tab:cov1}
    \vspace{-5pt}
\end{table*}

\begin{table*}[htbp]
    \centering
    \caption{Comparison of relevant baselines for average line coverage and failure detection in second benchmark.}
    \resizebox{\textwidth}{!}{
    \begin{tabular}{c||ccccc||ccccc}
        \hline
        \multirow{2}*{Web apps} & \multicolumn{5}{c||}{Average Line Coverage (\%)}  
                       & \multicolumn{5}{c}{Average number of failures (\#)}  \\
        \cline{2-11}
                      & Crawljax  & FEEDEX & FragGen & WebExplor & WebRELD  
                      & Crawljax  & FEEDEX & FragGen & WebExplor & WebRELD \\
        \hline
        Timeoff   & 19.96 \% & 18.72 \% & 19.96 \% & 17.89 \% & \textbf{44.08} \%  
                      & 1.0 & 0.8 & 1.0 & 0.0 & \textbf{7.2} \\
                      
        Realworld    &  55.40 \% & 55.40 \% &  55.40 \% &  55.40 \% & \textbf{81.30} \%  
                      & 2.0 & 2.0 & 2.0 & 2.0 & \textbf{5.4} \\
                      
        4gaBoards  & 61.89 \% & 59.80 \% & 61.59 \% & 59.73 \% & \textbf{63.17} \%  
                      & \textbf{2.0} & 0.0 & \textbf{2.0} & 0.0 & \textbf{2.0} \\
                      
        Parabank  & 18.20 \% & 17.00 \% & 18.00 \% & 18.60 \%   & \textbf{32.80} \%  
                      & 2.0 & 1.6 & 1.8 & 4.0 & \textbf{4.2} \\
                      
        Gadael    & 35.23 \% & 31.55 \% & \textbf{42.97} \% & 31.60 \%   & 41.06 \%  
                      & 2.0 & 1.0 & \textbf{5.0} & 1.0 & \textbf{5.0} \\
        
        Agilefant & 23.00 \% & 24.20 \% & 32.60 \% & 32.20 \%   & \textbf{34.80} \%  
                      & 4.0 & 3.2 & 4.6 & 4.0 & \textbf{5.2} \\
        \hline
        Average    & 35.61 \% & 34.45 \% & 38.42 \% & 35.90 \% & \textbf{49.54} \% 
                      & 2.2 & 1.4 & 2.7 & 1.8 & \textbf{4.8} \\
        \hline
    \end{tabular}}
    \label{tab:cov2}
\end{table*}

The left part of Table~\ref{tab:cov1} presents the code coverage results from the first benchmark, with the bold number indicating the best result. Our main findings are as follows: In the first benchmark, RL-based WebExplor outperforms Crawljax, FEEDEX, and FragGen in branch coverage. This is because WebExplor generates high-quality action sequences guided by the designed reward model. While Crawljax incrementally builds state machines to assist exploration, its state abstraction technique is ineffective, often missing critical pages that impact business logic. Unlike adaptive guidance in reinforcement learning, FEEDEX’s feedback effectiveness is constrained by fixed weights, limiting its adaptability across different applications. Fragment-based state abstraction enables FragGen to perform well in applications like Pagekit and Petclinic. However, without feedback guidance, FragGen only reproduces previously executed actions after exploration periods, limiting its ability to uncover new business logic. WebRLED achieves competitive code coverage, surpassing the state-of-the-art WebExplor by more than 8\% on average across all web applications in the first benchmark, as confirmed by the Mann-Whitney U test~\cite{korosteleva2013nonparametric} at a 0.05 confidence level.

The left part of Table~\ref{tab:cov2} compares the code coverage of WebRLED with baseline approaches for the second benchmark. WebRLED's coverage is more than 11\% higher than the other approaches. FragGen achieves higher coverage on Gadael by recording and executing actions on all visited states. In contrast, WebRLED does not record every action, which may miss some actions given limited testing time. However, WebRLED can reach deeper states by guiding rewards through appropriate action combinations, which FragGen struggled to do. We plan to extend WebRLED with this mechanism for a better exploration. Besides, we manually inspect the exploration history to analyze why WebRLED achieves higher coverage on most subjects (5/6). We find that Q-Learning struggles to replicate complex behaviors. For example, in the Dimeshift application, goal creation is realized through a pseudo-form composed of divs. The correct path to create a goal from the start page is as follows: Here $\rightarrow$ Goals $\rightarrow$ Create New $\rightarrow$ Sample Cash Wallet $\rightarrow$ Confirm and Save $\rightarrow$ Goal Details.

Multiple consecutive correct decisions are required to achieve this, and even one wrong step can disrupt the goal creation process. The adaptability of Q-Learning is limited by the large state/action space, making it difficult to learn an effective exploration strategy. In contrast, the DRL algorithm uses DNNs to memorize the correct sequence of actions. Guided by appropriate rewards, it reduces interference from irrelevant actions. As a result, WebRLED can learn complex behaviors and navigate deeper states in web applications.

\finding{RQ1}{WebRLED achieves higher code coverage than other baseline approaches on 12 selected open-source web applications. In particular, it achieves 11\% higher code coverage on six complex applications, demonstrating its ability to explore complex state space. Moreover, due to DNNs and designed rewards, WebRLED can effectively learn complex behaviors and handle large action space better than other approaches.}

\subsection{Failure Detection (RQ2)}

The right part of Table~\ref{tab:cov1} shows the average number of failures found for the first benchmark. WebRLED detected a higher average number of failures on most subjects (4/6) than WebExplor.

The right part of Table~\ref{tab:cov2} compares the number of failures discovered by WebRLED and baseline approaches for the second benchmark. WebRLED found 2.1 more failures on average, demonstrating stronger failure detection capabilities (i.e., calculated by Mann-Whitney U test~\cite{korosteleva2013nonparametric} at 0.05 confidence level). As WebRLED explores more states, it has more opportunities to find failures.

\label{failure_detection}
\begin{figure}[htbp]
  \centering
  \begin{subfigure}[b]{0.45\linewidth}
    \includegraphics[width=\textwidth]{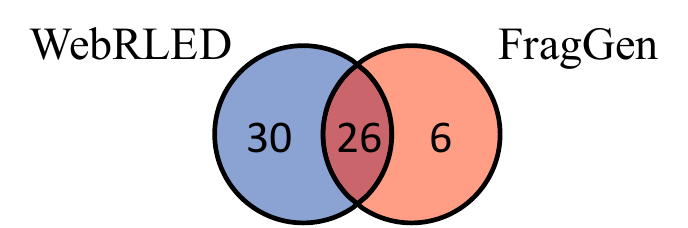}
    \caption{WebRLED vs FragGen}
    \label{fig:failures_comparison:FragGen}
  \end{subfigure}
  \hfill
  \begin{subfigure}[b]{0.45\linewidth}
    \includegraphics[width=\textwidth]{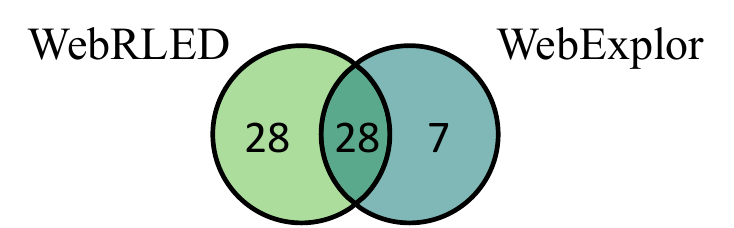}
    \caption{WebRLED vs WebExplor}
    \label{fig:failures_comparison:WebExplor}
  \end{subfigure}
  \caption{Pairwise comparison of unique failures across 12 applications.}
  \label{fig:failures_comparison}
  \vspace{-10pt}
\end{figure}

Figure~\ref{fig:failures_comparison} depicts the pairwise comparison of the unique failures captured by WebRLED, FragGen, and WebExplor across 12 web applications. The unique failures discovered by WebRLED are 5 times more than those found by FragGen and 4 times those found by WebExplor. This is because WebRLED explores more pages, increasing the chances of triggering failures. We also manually analyze some specific cases of discovered failures. For example, for Dimeshift application, login always fails (triggering internal server errors) after changing the password. After checking the code segment below, we find that this issue is caused by mistakenly assigning the new password to the login account.

\begin{lstlisting}
   function(req, res, next) {
   -   password = api.getParam(req, ``login'')
   +   password = api.getParam(req, ``password'')
       user.update({password: password,
       ...
\end{lstlisting}

For Timeoff application, an employee's available allowance should never be negative. Employees can submit one absence request when they have 1 available allowance. Once approved, the allowance is reduced to 0, preventing further requests. However, the allowance becomes negative because employees can submit a second request before the first is approved. Due to the lack of proper validation, the administrator can approve both requests consecutively, causing the available allowance to go negative.

\begin{lstlisting}
 employee.promise_allowance({year})
  .then(allowance_obj => Promise.resolve(
  [allowance_obj.number_of_days_available_in_allowance, employee]))
 )
\end{lstlisting}

\finding{RQ2}{Compared to other baselines (especially WebExplor), WebRLED detect more failures in most cases, as its strong exploration capabilities, which has more opportunities to trigger failures.}

\subsection{Ablation Experiment (RQ3)}
\begin{figure*}[htbp]
    \centering
    \includegraphics[width=\linewidth]{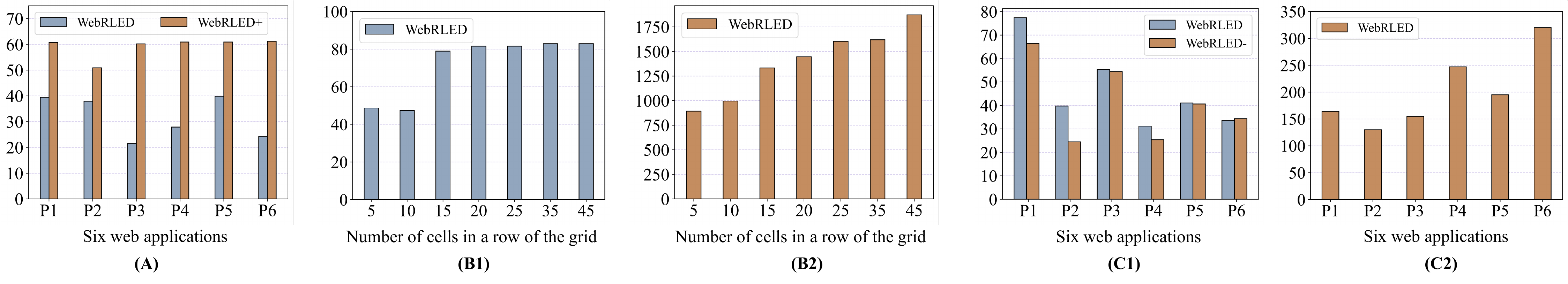}
    \caption{\textbf{(A):} Time (\textit{minutes}) consumed by WebRLED and \textit{WebRLED+} to achieve the same coverage across six different web applications. \textbf{(B1) \& (B2):} Coverage (\%) and time (\textit{seconds}) spent at the 35th episode for WebRLED with different number of cells in the grid on Phoenix App. \textbf{(C1):} Code coverage(\%) running WebRLED and \textit{WebRLED-} for one hour on selected subjects. \textbf{(C2):} The consumed training time (\textit{seconds}) of action discriminator by running WebRLED for 1 hour on selected subjects. \textbf{P1 to P6} correspond to the following web applications: Retroboard, Pagekit, Dimeshift, Parabank, Gadael and Agilefant.}
    \Description{\textbf{(A)} Time(\textit{minutes}) consumed by WebRLED and \textit{WebRLED+} to achieve the same coverage across six different web applications. \textbf{(B1)\&(B2)} Coverage(\%) and time(\textit{seconds}) spent at the 50th episode for WebRLED with different number of cells in the grid on Phoenix App. \textbf{(C1)} Code coverage(\%) running WebRLED and \textit{WebRLED-} for one hour on selected subjects. \textbf{(C2)} The consumed training time (\textit{seconds}) of action discriminator by running WebRLED for 1 hour on selected subjects. \textbf{P1 to P6} correspond to the following web applications: Retroboard, Pagekit, Dimeshift, Parabank, Gadael and Agilefant.}
    \label{fig:RQ3}
        \vspace{-10pt}
\end{figure*}

\textbf{Effectiveness of grid-based action value learning technique (\textit{WebRLED vs. WebRLED+})}. \textit{WebRLED+} implements list-based action space representation. Figure~\ref{fig:RQ3}.(A) compares the time cost for WebRLED and \textit{WebRLED+} to achieve the same level of code coverage. It can be seen that WebRLED takes significantly less time than \textit{WebRLED+}, and the time cost is reduced by 46\%. Based on the experimental result, we can conclude that the proposed grid-based  action value learning technique is effective and can significantly improve exploration efficiency.

We also evaluate the impact of the number of cells in the grid on exploration capability of WebRLED. The experiment is performed on Phoenix, and the grid is divided into $5\times5$, $10\times10$, $15\times15$, $20\times20$, $25\times25$, $35\times35$ and $45\times45$, respectively. Figure~\ref{fig:RQ3}.(B1) illustrates how the setting affects coverage. Figure~\ref{fig:RQ3}.(B2) shows the effect of the different setting on the time cost. It can be seen that increasing the number of cells affects both code coverage and the time. When the number of cells is less than 10, the effectiveness of WebRLED is comparable to the random-based approach, as the cells are too sparse to represent the value of the action. When the number of cells is greater than 10, the value of actions learned by WebRLED becomes more accurate. However, with increasing number of cells on the grid, the time required to select the action and update the DQN also increases. In particular, when the number of cells is greater than 20, the increase of the code coverage slows down, while the time increases significantly. To balance effectiveness and efficiency, we set the number of cells in the grid to a uniform 20. We found that this setting works well for the selected web apps in the experiments.

\begin{figure*}[htbp]
    \centering
    \includegraphics[width=\textwidth]{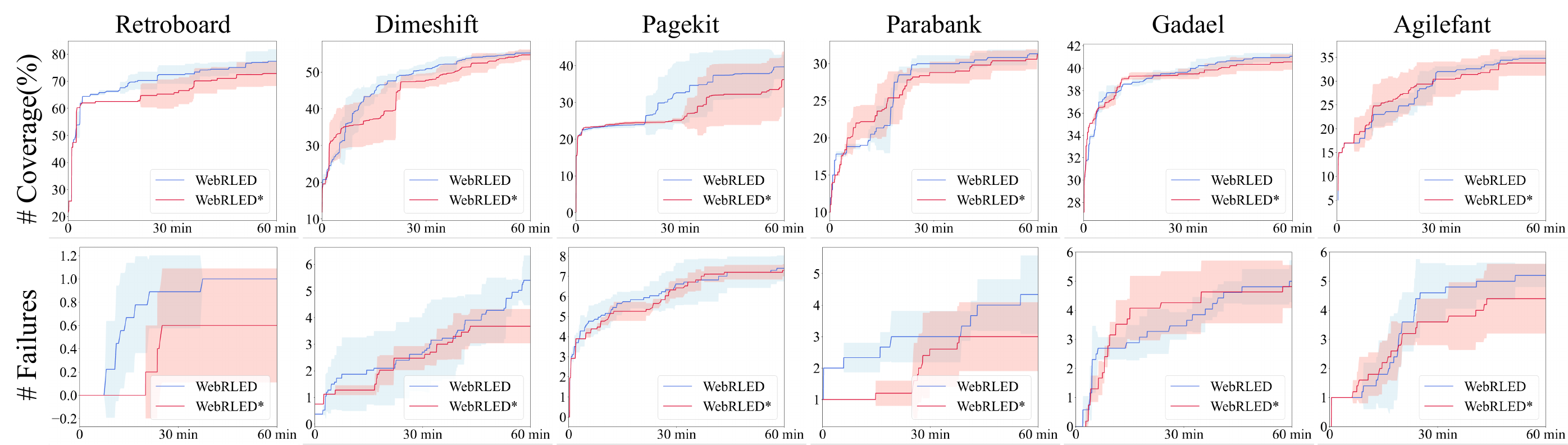}
    \caption{Evaluation of WebRLED compared to \textit{WebRLED*} in terms of code coverage (top) and average number of failures discovered (bottom). Solid lines represent the mean, and shading indicates the standard deviation.}
    \Description{Evaluation of WebRLED compared to \textit{WebRLED*} in terms of code coverage (top) and average number of failures discovered (bottom). Solid lines represent the mean, and shading indicates the standard deviation.}
    \label{fig:cov}
\end{figure*}

\textbf{Effectiveness and time cost of action discriminator (\textit{WebRLED vs. WebRLED-})}. \textit{WebRLED-} identifies actions without training an action discriminator. As shown in Figure~\ref{fig:RQ3}.(C1), WebRLED outperforms \textit{WebRLED-} in code coverage for most applications (5/6). The reason is that the action discriminator can recognize more actions on the page. However, \textit{WebRLED-} achieves more coverage than WebRLED in Agilefant because all actions can be identified using predefined heuristic rules, accounting for randomness. Additionally, we evaluate the training overhead of the action discriminator. As shown in Figure~\ref{fig:RQ3}.(C2), the average training time is about 4 minutes, striking a good balance between accuracy and time cost.

\textbf{Effectiveness of episodic reward model (\textit{WebRLED vs. WebRLED*})}. \textit{WebRLED*} just considers the global reward by removing the episodic reward. As shown in Figure~\ref{fig:cov}, WebRLED discovers more failures and achieves higher coverage compared to \textit{WebRLED*}. Initially, WebRLED explores slightly slower than \textit{WebRLED*}, especially on the subject Dimeshift, Pagekit and Agilefant. The reason is that under the guidance of episodic rewards, WebRLED explores states that have already been fully explored at the start of each episode, which slows down its coverage growth.
However, this also allows WebRLED to have the opportunity to discover new states later. In contrast, the reward of \textit{WebRLED*} degrades and cannot guide the exploration further in the later stage.

\finding{RQ3}{(1) The grid-based action value learning technique in WebRLED significantly outperforms the list-based technique, which reduces time by 46\% to achieve the same code coverage. (2) The action discriminator allows WebRLED to achieve higher coverage by triggering more actions. (3) The episodic reward model can guide the exploration continually, achieving higher coverage and discovering more failures.}

\subsection{Scalability (RQ4)}
To evaluate the effectiveness of WebRLED in real-world web applications, we chose the top 50 web applications based on the Alexa rank list~\cite{top50}. In total, WebRLED discovered 7,657 failures. We categorized them by source URL and discovered that 56.1\% originated from the applications under test, while 43.9\% came from third-party libraries. This indicates that most failures indeed exist in the original application and need to be tested using tools like WebRLED.

We manually deduplicated the found failures, identifying a total of 695 unique failures. We found that these failures include: (1) 106 JavaScript failures, such as syntax errors, semantic errors, and runtime errors. (2) 343 network-related failures, including resource loading failures and cross-domain failures. (3) 135 content security policy violations and other failures. Sometimes, a single failure can lead to multiple other failures. For instance, failing to load a critical JavaScript file can cause execution errors in dependent files. Furthermore, our analysis of HTTP status codes revealed that 45\% of the failures originate from the client side, while the rest come from the server. We also assess the severity of failures based on whether the failures can be observed on the page. There are 48 serious failures (e.g., web pages displaying 404 errors, thrown error message on the page).

\finding{RQ4}{WebRLED detected 695 unique failures in the top 50 most popular real-world web applications, further proving its effectiveness in failure detection.}

\section{Threats to Validity}
\textbf{Internal Threats.} The selection of web applications could be biased. To address it, most of the applications that we selected (9/12) come from prior research work~\cite{biagiola2019diversity,yandrapally2023carving}, which has been extensively used by existing studies~\cite{zheng2021automatic,chang2023reinforcement}. In addition, we also evaluate various real-world web applications, including active commercial applications. The main internal threat lies in the settings of parameter settings for all baseline tools. To mitigate the threat, we applied identical settings to all tools. This includes configuring the same login script, setting the same action waiting time, etc.

\textbf{External Threats.} To evaluate the scalability of WebRLED, we evaluate it on 12 open-source applications and 50 real-world applications, covering diverse technologies such as Java, PHP, and Node.js. Among them, 6 open-source applications are selected for their higher complexity. Randomness is one of the main threats in testing. To reduce this threat, We repeated each experiment five times and used non-parametric test to compare the results.

\textbf{Construct Validity.}
Like related work~\cite{zheng2021automatic,chang2023reinforcement}, WebRLED uses coverage and the number of failures as metrics to evaluate test effectiveness. In our experiments, we find that the real-time coverage in Splittypie and Pagekit decreased over time, which is not expected. This happens because the total number of tracked files increases as new web pages are discovered. To ensure a fair comparison, we recalculate coverage using a fixed number of files. Additionally, we categorize the discovered failures by type and severity.

\section{Related Work} \label{related_work}

\textbf{Model-Based Testing.} 
Model-based approaches~\cite{mesbah2012crawling,fard2013feedback,yandrapally2022fragment,amalfitano2014mobiguitar,amalfitano2012using,baek2016automated,gu2017aimdroid,lai2019goal,wang2020combodroid} construct a navigation model of web applications through static or dynamic analysis. Test cases generated from prior knowledge in the model can trigger complex business logic within the application. For instance, Crawljax~\cite{mesbah2012crawling} generates test cases by traversing dynamically constructed models. 
FRAGGEN\cite{yandrapally2022fragment} proposes a fragment-based state abstraction that accurately identifies states by representing them as fragments. Unlike the fragment-based approach, the grid-based approach of WebRLED is designed to address the action misalignment problem rather than state identification.
FEEDEX~\cite{fard2013feedback} uses DOM and path diversity to guide a crawler towards the generation of more accurate testing models. However, constructing and maintaining a comprehensive model for complex web applications is a challenge. Unlike these works, WebRLED can automatically learn the behavior of the application through DQN without constructing models.

\textbf{Systematic Strategies.} 
Systematic strategies~\cite{mao2016sapienz,dong2020time,biagiola2017search,biagiola2019diversity,su2017guided,artzi2008finding} use evolutionary algorithms or symbolic execution to generate inputs that target coverage scope. For example, Biagiola et al. employed search-based~\cite{biagiola2017search} techniques, diversification~\cite{biagiola2019diversity} of test events to generate test cases. Apollo~\cite{artzi2008finding} applies symbolic execution technique to generate specific inputs to cover hard-to-reach codes. However, generating specific inputs is not the primary concern of WebRLED currently, and other advanced input generation techniques can be integrated with WebRLED to improve effectiveness.

\textbf{Reinforcement Learning Based Testing.} Reinforcement learning has been widely used in Android testing~\cite{mariani2012autoblacktest,koroglu2018qbe,pan2020reinforcement,romdhana2022deep}. 
Q-Testing~\cite{pan2020reinforcement} uses deep learning to calculate rewards based on the similarity between states, but its backbone is tabular Q-Learning. 
ARES~\cite{romdhana2022deep} is the first Android testing work based on DRL, demonstrating the advantages of DRL over tabular RL in complex environments. However, due to the differing characteristics of Android and web applications, the challenges addressed by WebRLED and ARES are also different. Existing work in both the DRL and web domains includes WGE~\cite{liu2018reinforcement}, Dom-Q-Net~\cite{jia2019dom} and CC-Net~\cite{humphreys2022data}. These tools use DRL to complete well-defined tasks on the web. However, these techniques cannot be directly applied to solve web testing problems. In web testing, the first initiative to use reinforcement learning to generate test cases was WebExplor~\cite{zheng2021automatic}. It constructs an automaton based on curiosity-driven principles to provide high-level guidance for exploration, representing the state-of-the-art in current web testing technology. QExplore~\cite{sherin2023qexplore} is also a method based on tabular Q-Learning, which incorporates a novel text input generation technique. However, the backbone of them remains tabular Q-Learning, which struggles to solve the state explosion problem in complex web applications. UniRLTest~\cite{yu2022universally} is a cross-platform (Android and web) that relies on computer vision techniques to extract the states and actions of the screenshots. WebRLED focuses on testing web applications by retrieving the state and identifying actions through HTML documents.

\section{Conclusion} \label{conclusion}
In this paper, we propose WebRLED, a novel DRL-based approach for effective web testing. Specifically, we propose a grid-based action value learning technique that can dramatically improve exploration efficiency. A novel action discriminator is designed to identify more actions during exploration. We also design an adaptive reward model that considers the novelty of an explored state within an episode and global history, and can guide exploration continuously. Experiments show that WebRLED outperforms existing SOTA techniques. In the future, we plan to apply WebRLED to more real-world applications to evaluate its effectiveness in practice.

\bibliographystyle{ACM-Reference-Format}
\bibliography{sample-base}

\end{document}